\documentclass[fleqn,10pt]{wlscirep}
\usepackage[utf8]{inputenc}
\usepackage[T1]{fontenc}
\title{Observational Validation of Cosmic Ray Acceleration Hypothesis}
\author[1,*]{Anil Raghav}
\author[1]{Kalpesh Ghag}
\author[1]{Omkar Dhamane}
\author[2]{Zubair Shaikh}
\author[3]{Ankush Bhaskar}
\author[1,4]{Utsav Panchal}

\affil[1]{ Department of Physics, University of Mumbai, Mumbai, 400098, India}
\affil[2]{Indian Institute of Geomagnetism, Panvel, Navi Mumbai, India}
\affil[3]{Vikram Sarabhai Space Centre (VSSC), Indian Space Research Organisation (ISRO), Thiruvananthapuram, Kerala 695022, India}
\affil[4]{Department of Mathematics, Physics and Electrical Engineering, Northumbria University, Newcastle upon Tyne, NE1 8ST, UK}
\affil[*]{anil.raghav@physics.mu.ac.in}

\begin{abstract}
Despite centuries of rigorous theoretical and observational research, the origin and acceleration mechanism of Galactic Cosmic Rays (GCRs) remain a mystery. 
	In 1949, Fermi proposed a diffusive shock acceleration model that includes a prominent mechanism for GCR acceleration. However, observational evidence, on the other hand, remains elusive. Here,  we provided the first apparent verification of GCR acceleration at 1 AU using measurements from the CRIS instrument onboard the ACE spacecraft.
\end{abstract}
\begin{document}

\flushbottom
\maketitle
%
%
\thispagestyle{empty}
\section{\label{sec:1}Introduction}
	
	Victor Hess's balloon trip experiment in 1912 to explore the strength of ionizing radiation as a function of altitude marks the start of the cosmic rays study \cite{blackett1934CR,cronin2014spontaneous}. Since then, we've learned a lot about them, including their high energy range, composition, and their flux behavior as a function of energy \cite{nagano2000observations}; however, their source of origin remain a mystery \cite{friedlander2012century,sigl2001ultrahigh,diehl2009particle}. Recent discoveries suggest that the active galactic nuclei (galaxies with a supermassive core black hole) and gamma-ray bursts as probable sources of the highest-energy cosmic rays, whereas galactic sources include Type II supernovae, pulsars, and supernova remnants shock \cite{abraham2007correlation,abraham2008correlation,aharonian2019massive,sigl2001ultrahigh,ackermann2011cocoon,abeysekara2021hawc}.

The cosmic rays are either produced as a consequence of heavy particle disintegration \cite{hooper2006pierre}, or accelerated inside locations of strong magnetic fields \cite{havnes1971abundances,ostrowski2002mechanisms}. 
   Reported studies depict two basic types of cosmic ray acceleration mechanisms: diffusive acceleration and inductive (one-shot or direct) acceleration \cite{hillas1984origin,ptitsyna2010physical}. The Fermi first-order \cite{fermi1949origin} and second-order \cite{blandford1987particle,blandford1978particle} accelerations are prime examples of diffusive processes.  In contrast, acceleration by the large-scale electric fields is an example of inductive techniques \cite{schopper2002high}.
   
  Fermi's 1949 model described second-order acceleration processes for GCRs that relate the amount of energy gained by the particle's motion in the presence of randomly moving `magnetic mirrors (magnetic clouds).' As a result, if the magnetic mirror moves in the direction of the particle, the particle gains energy from reflection; However, if the mirrors move in the other direction, the particle loses energy.  In the case of cosmic rays energization, the magnetic mirror is a randomly moving interstellar magnetized cloud with a constant velocity.  According to Fermi, the chance of a head-on collision is higher than that of a head-tail collision; hence particles, on average will be accelerated. Here, the mean energy gain per bounce is proportional to the square of the mirror velocity, Thus, it is referred to as second-order Fermi acceleration \cite{achterberg1984stochastic}.  
   %

It is worth noting that this mechanism leads to a power‐law energy spectrum. However, there are some limitations to this process. The random velocities of interstellar clouds in the Galaxy are extremely low compared to the velocity of light. Therefore, the mean-free path for cosmic ray scattering in the interstellar medium is of the order of $0.1 ~pc$. Hence, the number of collisions would be limited to a few per year, resulting in a very small gain of energy by the particles \cite{palmer1982transport}. Thus, another particle acceleration process is proposed based on the strong astrophysical shock waves \cite{lee2012shock,petrosian2012stochastic}. This process is referred to as diffusive shock acceleration or first‐order Fermi acceleration as gain energy is proportional to the shock velocity, i.e., 
   \begin{equation}
    	\langle \frac{\Delta E}{E}\rangle = \frac{4}{3} \left(\frac{\Delta V}{C}\right)
   \end{equation}
    It also results in a power-law spectrum with an energy spectral index of 2 \cite{lichtenberg1980fermi, ahn2009energy}, i.e.,    
    \begin{equation}
    	N(E)dE \propto E^{-2} dE
    \end{equation}

  It is the most popular model that explains how supernova remnants accelerate protons \cite{ahlers2009cosmic}. The supernova remnant is thought to be an expanding spherical shell of material pushing out into the interstellar medium. The ejected  material  in supernova explosions has velocities of the order of $10^4~ km/s$, which is substantially faster than the sound and Alfv\'en speeds ($\sim 10 ~km/s$) in interstellar medium \cite{berezhko2003shock,armillotta2022cosmic}. This generates a shock wave at the shell front and propagates through a diffuse medium at a supersonic speed, which is accompanied by turbulent magnetic fields both at downstream and upstream \cite{niemiec2006cosmic}.  
    A charged particle, like a proton, bounces back and forth between these two downstream and upstream fields, passing through the shock front repeatedly and accumulating additional energy with each pass \cite{sarris1974effects, sagdeev1991collisionless}. Eventually, it will build enough energy to escape the magnetic fields and kick off into space like a high-energy cosmic ray \cite{hillas2005can,slane2015supernova,caprioli2012cosmic}.     
    
    Recent gamma-ray studies of supernova remnants are seriously testing our understanding of how particles accelerate at fast shocks \cite{hillas2005can,slane2015supernova,caprioli2012cosmic}. However, at high energies, several issues arose, e.g., how to explain the smooth continuation of the cosmic-ray spectrum far beyond $10^{14}~ eV$, the very low level of TeV gamma-ray emission from several supernova remnants, and the very low anisotropy of cosmic rays.
     In fact, the cosmic-ray spectrum needed to account for the observed $E^{-2.7}$ spectrum, which is steeper than the test-particle prediction of first-order Fermi acceleration, i.e., $E^{-2}$ \cite{meli2013active,caprioli2012cosmic}. As a result, the offered robust model should be either revised or further observational data should be found to validate it.

 \section{Hypothesis}
    
   In this article, we attempted to give observational evidence of low energy cosmic rays acceleration within the heliosphere at 1 AU in a framework of diffusive shock acceleration. The 11-year solar activity cycle has been seen for millennia through the variations in the Sun's appearance as measured by sunspot counts and sporadic transitory events such as flares, coronal mass ejections (CMEs), and so on \cite{vaquero2009sun}.    	
   	 Out of them, CMEs are massive eruptions of energy and magnetized plasma from the solar corona into the solar wind \cite{low1994magnetohydrodynamic,low2001coronal}.  During the maximum phase, the Sun emits roughly three to five CMEs daily, whereas just one or two CMEs per week during solar minima. Spacecraft's in-situ measurements show that CME average/ maximum velocities vary from 200 to 2,000 km/s, with an average speed of 489 km/s. Their higher relative speed over the ambient solar wind produces a shock wave and a turbulent sheath at the CME front, seen directly in the corona by coronagraphs and in in-situ measurements \cite{kilpua2017coronal,manchester2005coronal,rouillard2011relating,reiner2007coronal}. Consequently, many shock waves are generated and propagated in the heliosphere during the maximum solar phase. The ambient GCR particle has favorable conditions for the frequent back-and-forth reflections from the ICME shocks in the heliosphere. As a result,  ambient GCR particles may gain more energy with each reflection and eventually get acceleration. This is similar to supernova remnants' conditions; however, in our case, it applies to lower-energy GCRs.

\section{Data, Methods, and Observations}
   
\begin{figure*}
  	\centering
  	\includegraphics[width=\textwidth]{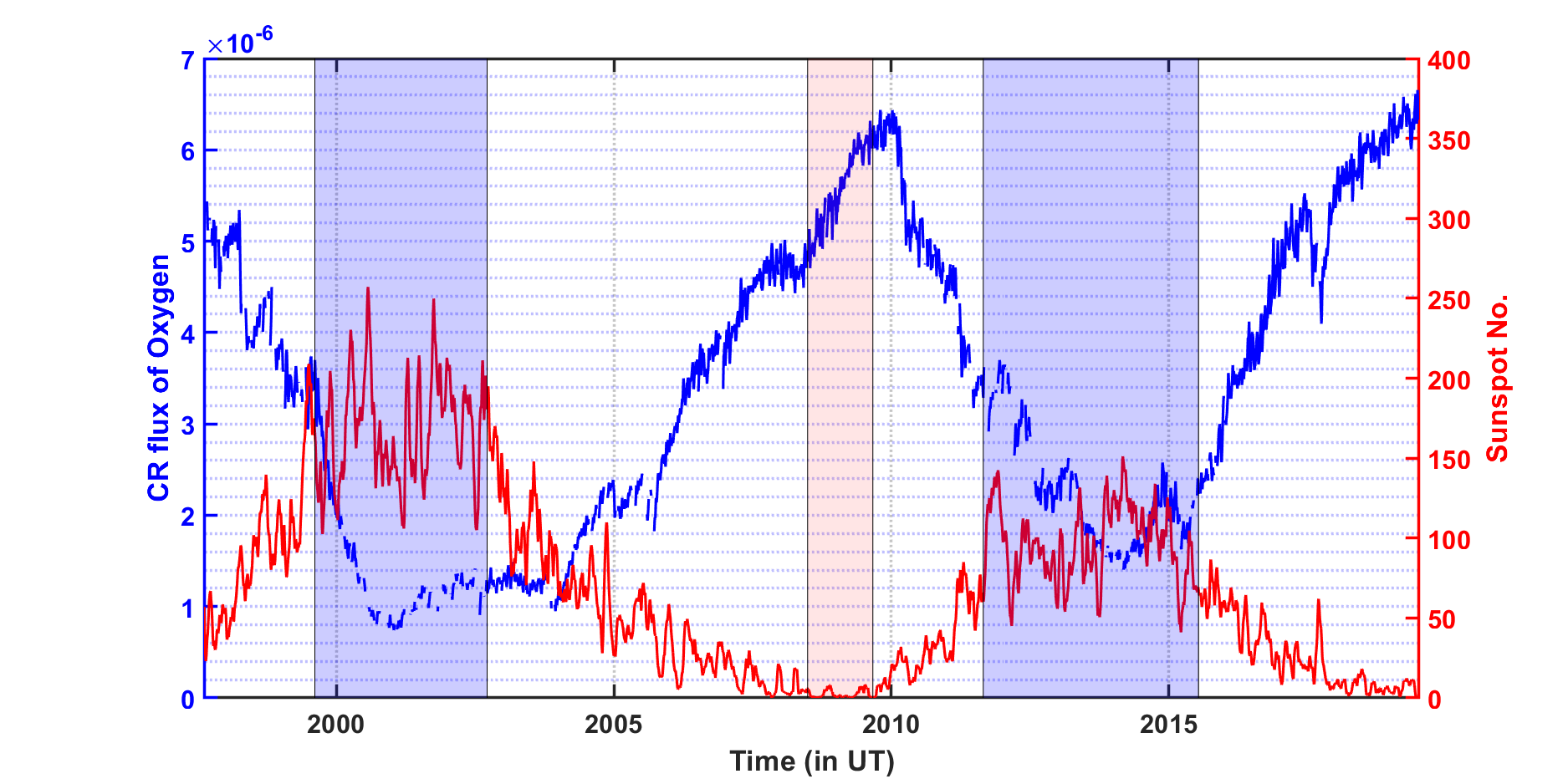}
  	\caption{Blue colored line indicates the CR flux of Oxygen. The red colored line shows the variation of solar activity, i.e., daily sunspot numbers. The blue shaded region indicates the solar maxima of both solar cycle 23 $\&$ cycle 24. The light red shaded region indicates the solar minima of both solar cycles 23 $\&$ 24.}
  	\label{Solar_cycle}
  \end{figure*}

 Figure \ref{Solar_cycle} shows the time variation of solar activity cycles (\url{https://cdaweb.gsfc.nasa.gov/cgi-bin/eval2.cgi}) along with the GCR variation using data recorded by the Cosmic Ray Isotope Spectrometer (CRIS) \cite{stone1998cosmic} onboard the Advanced Composition Explorer (ACE) \cite{stone1998advanced} satellite at 1 AU. Here, we utilized flux variations of the oxygen element as a proxy for GCRs. The ACE's CRIS instrument has a geometrical acceptance of $\sim 250~ cm^{-2} sr^{-1}$ which measures high-precision cosmic-ray nuclei mass, nuclear charge, and incident energy using the multiple $-dE/dx$ versus $E_T$ spectrometer. It measures the ion's elemental and isotopic composition for He to Ni ($2 \leq  Z \leq 28$) at energies ranging from 50 to 550 MeV/nucleon with elemental fluxes in 7 energy window intervals for each element  \cite{stone1998cosmic}. The archival data of CRIS is available at \url{https://spdf.gsfc.nasa.gov/}. We observed the significant elemental fluxes for $ C,~ N, ~O, ~Ne,~ Mg,~ Si~ \& ~Fe$; thus, we used their integral count data with 1 hr resolution for further analysis.


To investigate the acceleration process, we divide the data set into three separate regions (see Figure \ref{Solar_cycle}). The blue-shaded region indicates the maximum activity phase, whereas the red-shaded region indicates the minimum activity phase.  The selected boundaries used in the analysis are from 23 August 1999 to 20 September 2002 and 28 August 2011 to 15 July 2015 for a solar maximum of solar cycles 23 and 24, respectively. Similarly, the 28 June 2008 to 01 September 2009  period is used as solar minimum. 
  
  \begin{figure*}
	\centering
	\includegraphics[width=\textwidth]{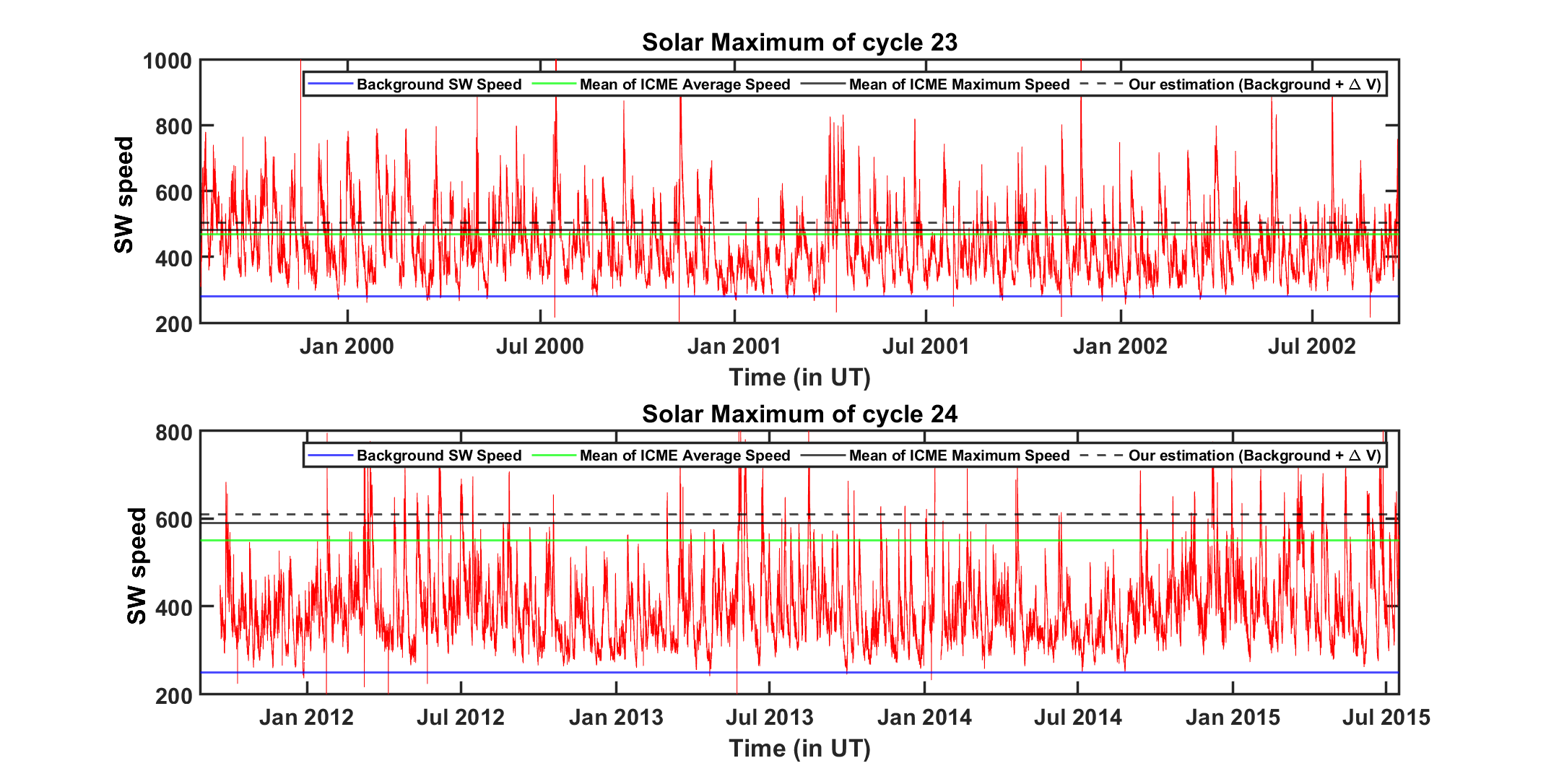}
	\caption{Top panel shows the solar wind (SW) speed variation for the solar maximum of cycle 23. The bottom panel shows the solar wind speed for a solar maximum of cycle 24. A solid blue line indicates the ambient solar wind speed. The dotted black line shows the solar wind's average value over the maximum solar period. The green line indicates the average speed of ICMEs for the maximum solar cycle. A solid black line illustrates the average of maximum solar wind in ICMEs over the solar maxima.   }
	\label{fig:solar_wind}
\end{figure*}

\begin{figure*}
    	\centering
    	\includegraphics[width=\textwidth]{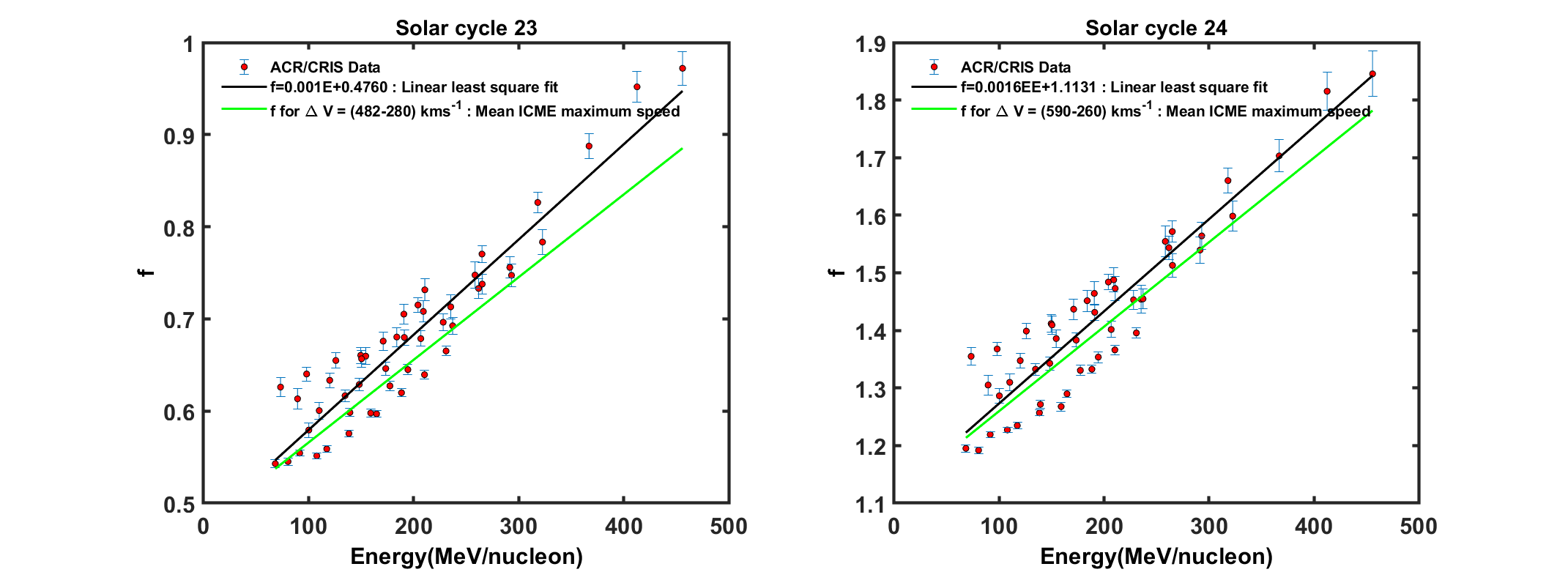}
    	\caption{The left and right plots show the linear relation between $f $ and energy for solar cycles 23 and 24, respectively.  The red dot shows the observational data of CRIS, and the solid blue line shows the linear fit.}
    	\label{ratio}
 \end{figure*}
    
   \begin{table*}
   \centering
\begin{tabular}{|c|c|c|}
\hline
\textbf{Speeds}                  & \textbf{For solar cycle 23 (km/s)} & \textbf{For solar cycle 24 (km/s)} \\ \hline
$\Delta V$ (by calculation)              & 225                                & 360                                \\ \hline
Background speed                 & 280                                & 250                                \\ \hline
Our estimation (Background + $\Delta V$) & 505                                & 610                                \\ \hline
Mean ICME max speed              & 482                                & 590                                \\ \hline
Mean ICME speed                  & 426                                & 551                                \\ \hline
\end{tabular}
\caption{The estimation of solar wind speed with distinct conditions are included here.}
\end{table*}

 \section{Discussion}
  Generally, a combination of four major physical processes controls GCR transport in the heliosphere, e.g., (1) Diffusion: caused by magnetic field abnormalities, (2) Convection: caused by solar wind plasma flow, (3) Guiding-centre drifts: such as gradient and curvature drifts, and (4) Energy change: caused by background plasma expansion and compression. The number of CMEs is less during solar minimum, implying a weaker interplanetary magnetic field. Thus high GCRs (maximum) flux is measured at each energy window of CRIS. We treat these counts as base values and use them for normalization. We observed many CMEs and their respective shock fronts during the maximum solar phase. The magnetic field is stronger and more complex during solar maximum. Therefore, GCR's interaction with them causes more scattering of the particles, resulting in lower GCR intensity in the heliosphere. The supporting corroborative visible anti-correlation is also evident in Figure \ref{Solar_cycle}.

   
  During each solar cycle's selected maximum (minimum) period, we summed over the integral counts measured in each energy window of CRIS.  Let's assume that $N_{min}(E_1)$ denotes the total number of counts collected in the $E_1$ energy window during the minimum phase. Therefore, the $N_{min}(E_r)$ will be associated with the $E_r$ energy window, and so on.   Based on the diffusive shock model, we anticipate an acceleration of GCR particles via their multiple to-and-fro reflections through CME sub-structures during solar maximum. Therefore, we assume that $N_{max}(E_1)$ is the number of counts detected in the $E_1$ energy window during the maximum phase. As a result, the $N_{max}(E_r)$ count is linked to the $E_r$ energy window, and so on. These counts are the sum of a base count (similar to counts during solar minimum) and a count rise caused by acceleration. Furthermore, the overall cosmic ray flux at the solar minimum is greater than the maximum solar period, driven by different heliospheric irregularities (see anti-correlation in Figure \ref{Solar_cycle}). As a result, we estimate the normalized CR flux for each energy window as
   \begin{equation}
  	f = \frac{N_{max}(E_r)}{N_{min}(E_r)}
  \end{equation}

In the literature, particle flux enhancement is used as a proxy for particle energization/acceleration \cite{zaharia2000particle,klein2010energetic,rice2003particle}. We hypothesized that the diffusive shock mechanism elevated the energy of particles moving in the heliosphere during the Sun's high-activity period.
Due to the energization process, a higher energy window may now detect the particle flux from its immediate lower energy window, contributing as an enhancement. Thus, considering the Fermi acceleration, we expect ratio $f$ is increasing with respect to the increase in E, i.e., midpoint energy for a given energy window of CRIS. Here, we assumed that the other processes like diffusion, convection, or guiding center drift contributed equally to the flux measured during solar maximum and minimum periods. Moreover, their contribution in determining ratio f should be canceled; thus, we neglect their effect in $f$.  As a result, we suggest a ratio $f$ is a suitable proxy for particle energy gain. i.e., 
  $f \approx \Delta E$ \cite{zaharia2000particle,klein2010energetic,rice2003particle}.

  Substituting in equation 1, we have 
\begin{equation}
	\frac{f}{E_r}	\approx  \frac{4}{3} \left(\frac{\Delta V}{C}\right)
\end{equation}
  After rearranging the terms, we have 
  \begin{equation}
  f \approx \frac{4}{3} \left(\frac{\Delta V}{C}\right)~~ E_r
  \end{equation}
  
  %
  %
  %
   
   %

  

  
 
where `$\Delta V$' is the excess speed of shock-front (ICME) over ambient solar wind speed, and `$C$' is the speed of light. Moreover, $f$ is obtained empirically by dividing maximum phase GCR counts by minimum phase GCR counts for each defined energy window of the CRIS instrument.

Figure \ref{ratio} demonstrates the variation of $f$ for the midpoint energy of each given energy window of the CRIS instrument for different particles. The scatter plot clearly shows a good correlation and regression between them. Furthermore, we perform linear least squares fitting (regression analysis) to the scattered data, where the slope of the fitting line provides the information of $\frac{4}{3} \left(\frac{\Delta V}{C}\right)$, which is equal to $\sim 0.001$ and $\sim 0.0016$ for solar cycles 23 \& 24 respectively.   Using this relation, we calculated the relative excess speed of the interplanetary shock of CMEs over ambient solar wind speed ($\Delta V$) as $\sim 225 $ km/s and $\sim 360$ km/s for solar cycles 23 and 24, respectively.

Moreover, to validate the above resultant ICME shock speeds derived from our model, we have used additional data-set  from SWEPAM (Solar Wind Electron Proton and Alpha Monitor) instrument onboard the ACE spacecraft. The data has a time resolution of 1 hour and is available at \url{www.srl.caltech.edu}.  Figure \ref{fig:solar_wind}  shows the temporal variation of solar wind speed during the maximum phase of solar cycle 23 (top panel) and 24 (bottom panel). The horizontal blue (black) line shows the solar wind's background (mean value) respectively in both panels. The background solar wind speeds are $280 km/s$ and $250 km/s$ for solar cycles 23 and 24, respectively, shown with a blue horizontal line in \ref{fig:solar_wind}. In addition to this, we have also used the ICME catalog available at  \url{http://www.srl.caltech.edu/ACE/ASC/DATA/level3/icmetable2.htm}. Further, we estimate the mean maximum plasma speed during the ICME transit period for the selected maximum phase of the solar cycle 23 and 24, represented by the green line in Figure \ref{fig:solar_wind} respectively. The mean of ICME maximum speeds are $482~km/s$ and   $590~km/s$ for solar cycles 23 and 24, respectively.

So, we estimated $\Delta V$ by subtracting the background solar wind speed from the mean of ICME maximum speed. Using the above definition, the estimated $\Delta V$ is $202~km/s$ and $330~km/s$ for solar cycles 23 and 24, respectively.  Moreover, the estimated $\Delta V$ from the slope of linear fitting to the CRIS dataset is $225~km/s$ and $360~km/s$, respectively.   Surprisingly, as expected, our estimated $\Delta V$ from the fitted slope closely corroborates with the mean of ICME maximum speed for solar cycles 23 and 24, confirming our hypothesis.

The Diffusive shock acceleration  (DSA)  theory can be used to estimate the upper limit of energy produced ($E_{max}$) during the process. By following Achterberg (2000),\cite{achterberg2000particle}  we get  $ E_{max} = ZeB\beta_{s}R_s$ . Where $Z$ is the atomic number, $e$ is the electron charge, $B$ is a magnetic field, $\beta_{s}$ is the ratio of shock velocity and speed of particles (which is $\sim$ speed of light for high energy particles), $R_s$ is the size of the source. Note that the near-Earth IP shock scale is almost 1AU; however, as it propagates into interplanetary space, we expect it to increase. Thus we have considered  $R_s \sim 10 AU$ which is more realistic. Moreover, the  IP  shock compresses the IMF and enhances it, which can give rise to $ 10s$ of nT field around it. Thus, we can consider realistic IMF B $\sim 10$ nT.  Considering the acceleration scenario for protons in IP shock and typical IP shock parameters [IMF B $\sim 10$ nT, $\beta_{s} \sim 10^{-3}$, $R_s \sim 10 AU$], we find $ E_{max} \sim 10 GeV$ which is sufficient energy and falls in the observed energy range of particles.

In summary, our finding confirms the applicability diffusive shock model hypothesis to GCR acceleration within the framework of the heliosphere. It implies that the cosmic ray acceleration via the diffusive shock model is also possible in supernovae remnants and stellar activities. In addition, the exponent value derived by the model, i.e., $-2$, is consistent with the observations of GCR variations in interplanetary space. Furthermore, it is also observed that  high energy CR flux shows steeper exponents, i.e., $-2.7$. It implies that cosmic ray flux reaching a particular observer point is of lower energy compared to the shallower exponent $-2$. This might be because cosmic rays lose their energy while propagating in the heliosphere before reaching the observer, which decreases flux. Moreover, one should not overlook the charged particle's energy loss due to various radiation loss mechanisms like bremsstrahlung, cyclotron, synchrotron, etc.    

	


\bibliography{CR}

\end{document}